\documentclass[letterpaper,twocolumn,10pt]{article}
\usepackage{usenix2019_v3}
\usepackage{amsmath,amssymb,amsfonts}
\usepackage{graphicx}
\usepackage{textcomp}
\usepackage{xcolor}
\def\BibTeX{{\rm B\kern-.05em{\sc i\kern-.025em b}\kern-.08em
    T\kern-.1667em\lower.7ex\hbox{E}\kern-.125emX}}
\usepackage{framed}
\usepackage{minibox}
\usepackage{amsmath,amssymb,amsfonts}
\usepackage{graphicx}
\usepackage{xcolor}
\usepackage[T1]{fontenc}
\usepackage[ruled,algosection,noend,noline]{algorithm2e}
\usepackage{multirow,array}
\SetMathAlphabet{\mathcal}{normal}{OMS}{lmsy}{m}{n}
\SetMathAlphabet{\mathcal}{bold}{OMS}{lmsy}{m}{n}
\SetSymbolFont{largesymbols}{normal}{OMX}{lmex}{m}{n}
\SetSymbolFont{largesymbols}{bold}{OMX}{lmex}{m}{n}
\DeclareMathAlphabet{\mathsc}{OT1}{cmr}{m}{sc}

\usepackage{enumitem}
\usepackage[all]{xy}
\usepackage{booktabs}
\usepackage{multirow}
\usepackage{url}
\usepackage{xspace}
\usepackage[group-separator={,}]{siunitx}
\usepackage{epigraph}
\usepackage{dsfont}
\usepackage{todonotes}

\newenvironment{denseitemize}{
\begin{itemize}
    \setlength{\itemsep}{1pt}
    \setlength{\parskip}{0pt}
    \setlength{\parsep}{0pt}
}{\end{itemize}}

\newcommand\bi{\begin{denseitemize}}
\newcommand\ei{\end{denseitemize}}
\newcommand\ben{\begin{enumerate}}
\newcommand\een{\end{enumerate}}
\newcommand{\trm}[1]{\textrm{#1}}

\newcommand*\dash{\ifvmode\quitvmode\else\unskip\kern.16667em\fi---%
\hskip.16667em\relax}

\usepackage{adjustbox}
\newcounter{countgap}
\definecolor{verylightgray}{gray}{0.90}
\newcounter{countinsight}
\definecolor{verylightgray}{gray}{0.90}
\newcounter{countdef}
\definecolor{verylightgray}{gray}{0.90}
\newcounter{counttechnique}
\definecolor{verylightgray}{gray}{0.90}
\newcommand{\strategy}{\sigma}
\newcommand{\strategyset}{\mathcal{S}}

\newcommand\precede\preccurlyeq

\newcommand{\adv}{\mathbf{Adv}}

\newcommand\saraha[1]{}
\newcommand\alex[1]{}
\begin{document}

\title{\Large \bf SoK: Tools for Game Theoretic Models of Security for Cryptocurrencies}

%for single author (just remove % characters)
\author{
{\rm Sarah Azouvi}\\
Protocol Labs\\ University College London
\and
{\rm Alexander Hicks}\\
University College London
% copy the following lines to add more authors
% \and
% {\rm Name}\\
%Name Institution
} % end author

\maketitle% Title
% \begin{center}
%   \Large\bfseries\boldmath
%   SoK: Tools for Game Theoretic Models of Security for Cryptocurrencies
% \end{center}

%!TEX root = main.tex
\begin{abstract}
Cryptocurrencies have garnered much attention in recent years, both from the academic community and industry.
One interesting aspect of cryptocurrencies is their explicit consideration of incentives at the protocol level, which has motivated a large body of work, yet many open problems still exist and current systems rarely deal with incentive related problems well. 
This issue arises due to the gap between Cryptography and Distributed Systems security, which deals with traditional security problems that ignore the explicit consideration of incentives, and Game Theory, which deals best with situations involving incentives.
With this work, we offer a systematization of the work that relates to this problem, considering papers that blend Game Theory with Cryptography or Distributed systems. This gives an overview of the available tools, and we look at their (potential) use in practice, in the context of existing blockchain based systems that have been proposed or implemented. %We conclude with \dots
\end{abstract}

%!TEX root = main.tex
\section{Introduction}
Since the deployment of Bitcoin in 2009, cryptocurrencies have garnered much attention from both academia and industry.
Many challenges in this area have since been recognized, from privacy and scalability to governance and economics.
In particular, the explicit consideration of incentives in the protocol design of cryptocurrencies (or ``cryptoeconomics'') has become an important topic.

The importance of economic considerations in security has been acknowledged since work by Anderson~\cite{anderson2001information,anderson2006economics}, who recognized that many security failures could be explained by applying established ideas from Game Theory (GT) and Economics.
However, the incentives at play tend to be external to the system design, and sometimes implicit, leading to failures when the the intended use of systems is misaligned with the incentives of users.

Cryptocurrencies, on the other hand, explicitly define some incentives in the design of their system, for example in the form of mining rewards, suggesting that they could be properly aligned and avoid traditional failures.
Unfortunately, many attacks related to incentives have nonetheless been found for many cryptocurrencies~\cite{selfishmining,verifierdilemma,minersdilemma}, due to the use of lacking models.
While many papers aim to consider both standard security and game theoretic guarantees, the vast majority end up considering them separately despite their relation in practice.

Here, we consider the ways in which models in Cryptography and Distributed Systems (DS) can explicitly consider game theoretic properties and incorporated into a system, looking at requirements based on existing cryptocurrencies.

\subsubsection*{Methodology}
As we are covering a topic that incorporates many different fields coming up with an extensive list of papers would have been quite challenging, and would lead to an output of much greater length.
In order to pick a representative subset of papers, we started by looking at existing surveys on the topic of Game Theory and Security~\cite{linial-survey,katzbridging,halpern-survey,beyond,shoham-survey,tim-survey}, as well as specific book chapters on the topic e.g., Chapter $8$ in the book by Nisan et al.~\cite{nisan2007algorithmic}.
We then looked at work published in popular Cryptography venues (e.g.,  IACR conferences), Security conferences (e.g., FC), as well as Distributed System venues (e.g., PODC) and interdisciplinary venues (e.g., WEIS, ACM Economics and Computation, P2PECON, GameSec), looking specifically for papers that cover both Game Theory and Security.

Different papers sometimes present different definitions for similar concepts, so we do not always include all these definitions.
We also omit to present work on security and game theory that does not directly relate to what we discuss, e.g., the 
body of work by Tambe et al.~\cite{tambe2011security} about physical security and allocation of limited security resources,
or by Grossklags et al.~\cite{grossklags2008secure} about security investments.
Instead, we focus on some specific models of interest that we think are more appropriate, and match these with problems in the security of cryptocurrencies.

\subsubsection*{Our contributions}
Our goal is to give an overview of the intersection of the three fields that are essential to the design of cryptocurrencies: Cryptography, Distributed Systems and Game Theory.
Our contribution is an analysis of existing work that proposes solutions to this problem.
Our analysis highlights new concepts introduced by these papers, as well as deficiencies.
We do this in the context of security requirements that we formulate, arguing that they address deficiencies in existing security models that fail to cover all aspects of a decentralized cryptocurrency.
We also discuss open challenges and how they could be addressed.

Section~\ref{sec:prelim} introduces Game Theory and cryptocurrencies, and discusses security in the context of a decentralized system.
In Section~\ref{sec:cryptoandgt} we then look at the intersection of Cryptography and Game Theory, followed by the intersection of DS and Game Theory in Section~\ref{sec:dsgt}.
%This allows us to review the low level building blocks of existing solutions, which we put in the context of our security notions and requirements.
We then look at how these results are used in Section~\ref{sec:blockchains}, where we look at proposed systems and their failures, tracing back to deficiencies identified in the two previous sections.
In Section~\ref{sec:discussion} we give a comparative table of some of the concepts presented in this paper and how they could be used in blockchain research,
we discuss the open challenges posed by failures that are observed, and how they could be addressed.
Finally, Appendix~\ref{glossary} collects formal definitions for the concepts presented in the paper.

% \saraha{Vlad's definition of cryptoeconomics:
% ``A formal discipline that studies protocols that govern the production,
% distribution and consumption of goods and services in a decentralized digital
% economy. Cryptoeconomics is a practical science that focuses on the design
% and characterization of these protocols.''}

%!TEX root = main.tex
\section{Related work}
The work that most closely resembles ours are the previous surveys
bridging
Computer Science and Game Theory~\cite{linial-survey,katzbridging,halpern-survey,beyond,shoham-survey,tim-survey}.
They were of great inspiration
for this work, but they are quite outdated (dating back to 2002, 2005, 2007, 2008, 2010) given the recent output of research tied to cryptocurrencies and other blockchain based systems.
% This field indeed gained new meaningful applications with the apparition of Bitcoin and
% some new works have been done after these surveys that are worth of attention.

On the topic of blockchains, many SoK papers and surveys exist that cover consensus protocols and security~\cite{bano2017consensus,garaysok,wang2018survey,stifteragreement,7163021,cachin2017blockchains,liu2019survey}.
These are very different from our work as we present general concepts and definitions related to designing decentralized systems with incentives.
In particular, many concepts presented in this paper were not introduced in the context of consensus, but rather in the context of secure multiparty computation (MPC) or other problems tied to distributed systems.
% We present how previous work
% at the intersection of GT and Computer Science could be used in the blockchain space.
Most of the work presented in this SoK does not directly concern blockchains, which is the motivation behind this work.

%!TEX root = main.tex
\section{Background}
\label{sec:prelim}

In this section we briefly introduce game theoretic tools that are mentioned throughout the paper such as solutions concepts and mechanism design (MD).
For a complete introduction to Game Theory, the reader is invited to look at any of the books  (or other resources) on the topic~\cite{gametheoryowen,jackson}.
For the sake of exposition we omit to cover concepts that are relevant e.g., correlated equilibria, Pareto efficiency and the single deviation test, as these do not appear in the papers we mention.
We also discuss the interface between Game Theory and Cryptography and Security in practice, and briefly introduce Distributed Systems and cryptocurrencies.

% Here we give some background on different concepts
% of Game Theory (GT), Cryptography and Distributed Systems (DS).
% We refer the reader to ... for a comprehensive litterature review.

% \begin{itemize}
% \item solution concepts in GT like NE, strategyproofness, revelation principle, (sequential equilibrium,)
% social welfare maximization, individual rationality use \url{https://www.cs.cmu.edu/~sandholm/cs15-892F15/Rahul-thesis.pdf}
% \item safety liveness in DS
% \item UC and game proof in crypto
% \end{itemize} 

\subsubsection{Game Theory and Mechanism Design}
% \subsubsection{Games and solution concepts}
% We refer the reader to~\cite{jackson} for a more complete background
% on Game Theory and Mechanism Design.
A game is defined by a set of players and a set of actions
for each player.
A strategy for player is defined as a function from its local state to actions. 
%We denote $\strategyset_i$ the set of all possible strategy for player $i$.
%We denote $\vec{\strategy}=(\strategy_1,\cdots,\strategy_n)$ the joint strategy of all players
%and $\vec{\strategy}_{-i}=(\strategy_1,\cdots,\strategy_{i-1},\strategy_{i+1},\strategy_n)$.
%We denote $u_i(\vec{\strategy})$ player's $i$ utility when $\vec{\strategy}$ is played.
%Players may be of different types, which are denoted $\theta_i$ for player $i$.

Game Theory uses solution concepts in order to predict the outcome of a game, the most well known is the \textit{Nash equilibirum} (NE).
A strategy is a Nash equilibrium if given that all the other players follow this strategy, player $i$ is better off playing it as well.
It is often unrealistic to assume that players have complete information about the game they are in.
A game where players do not always know what has taken place earlier in the game is said to have \textit{imperfect information}.
In the case where players do not know the \textit{type} of the other players, which determines their utility function, the game is said to have \textit{incomplete information}.

%Imperfect and incomplete extensive form games can be related by having Nature make the first move in the game and randomly assign types to players, turning an incomplete game into an imperfect game as different outcomes of Nature's initial move are possible.
% The players then have a probability distribution over the types of other players. %(Formally speaking, the probability distribution is over the set of states of Nature, which are mapped to player types.)
%The beliefs of players are expressed as conditional probabilities of a history in an information set, given the information set.
The players then have a probability distribution over the types of other players. Their beliefs are expressed as conditional probabilities based on the information they have, which they can update using Bayes' theorem when they gain new information, leading games of this form to be called \textit{Bayesian games}.
%A collection of beliefs is called a \textit{belief system}, which paired with a strategy profile is an \textit{assessment}.

%Without perfect information, some solution concepts must also be abandoned as players cannot always tell which subgame they are in and thus a SPE is no longer possible.
Players now also think about expected payoffs, so the standard definition of a NE is no longer ideal but we can define a \textit{Bayesian Nash equilibrium} (BNE) analogously by replacing utilities with expected utilities, although we still refer to them simply as utilities.

While Game Theory is typically about understanding the behavior of players in a given game, systems are usually designed and implemented with a goal in mind e.g., preventing double spending in cryptocurrencies.
To achieve this, our goals can be expressed as a social choice function (SCF), a function that given the preference (or types) of all players outputs an outcome.
For example, in a voting system, given all the ranked preferences of voters, a SCF will
choose a candidate.
% We can also define a social welfare function (SWF) as a function that given the preference of all players outputs a global preference. For example, given all the ranked preferences of voters, a SWF will
% output a ranked list of candidates.

Once we have a target outcome in mind, the idea is to make sure that %it happens.
the incentives are designed in a way such that selfish players reach this outcome.
In some ways, this can be thought of as reversing the basic idea of Game Theory, designing a game that leads to a specific outcome.

%To do this, we use a mechanism that maps the action profile of players to a distribution over outcomes.
%A mechanism $M$ is then said to implement a SCF $f$  if $M(C_M(\vec{u}))=f(\vec{u})$, where
%$\vec{u}$ is the vector of all of the players utility functions and
%$C_M(\vec{u})$ represents all strategy vectors that could reasonably result from selfish behavior.
%Informally, the previous equality means that whatever selfish strategy players
%choose, the outcome of the game will correspond to the SCF.
%The solution concept is supposed to reflect reality, in the sense that when it holds, players' selfish strategies lead to the desired outcome.
For example in a voting system we would like to design a system where given all the preferences
of the players, the one chosen by the SCF is elected, one way to do this is to incentivize players to
report their truthful preferences.
A mechanism can also be viewed as a protocol, with the corresponding game being thought of as having the protocol as the recommended strategy, and deviations from the protocol as other possible strategies.

It must be pointed out that doing this in practice is not always easy, as experimental Game theory reveals. Gneezy and Rustichini~\cite{gneezy2000fine} looked at the effects of implementing fines at a nursery in order to reduce the rate at which parents collected their child late. Intuitively, this should motivate parents to arrive on time but parents instead interpreted the fine as a way of paying for extra childcare, and started coming even later.
% \alex{Could also mention the field of experimental game theory that looks at confirming theoretical results in experiments, as this is something that's often neglected in discussions.}
% This was done by punishing late parents with a fine, which intuitively should motivate parents to arrive on time. Instead, parents interpreted the fine as a way of paying for extra childcare, and started coming even later.
Furthermore, once the fine was removed as it was counter productive, the parents behavior did not revert back so the system had been irreversibly damaged.

A very important result in MD is the revelation principle %~\cite{jackson}
that states that any social choice function that can be implemented
by any mechanism can be implemented by a direct truthful mechanism. A mechanism is direct if players
need only reveal their type/utility function to the designer of the game and truthful if players'
best strategy is to reveal their true type.

% \alex{Do we talk about transferable utilities much later on?}
% \paragraph*{Transfer function}
% In order to provide the incentives necessary to make efficient choices, it may be
% necessary to tax or subsidize various individuals
% (e.g., add a financial punishment).
% Adjustments are made by a transfer function $t:\Theta\rightarrow\mathbb{R}^n$,
% where $t_i(\theta)$ represents the payment that player $i$ receives (if $t_i(\theta)<0$)
% or make (if $t_i(\theta)<0$).
% One example of transferable utility function is quasi-linear utility function which has the form:
% %$u_i(\vec{\theta},\theta_i,d,t)=v_i(d(\vec{\theta}),\theta_i)+t_i(\vec{\theta})$
% $u_i(\vec{\theta},t)=v_i(\vec{\theta})+t_i(\vec{\theta})$
% where $v_i$ is the agent's valuation for the outcome.

% Game Theory often assumes the existence of a trusted mediator or center
% to enforce the game's rules.
% This is unrealistic or hard to implement in practice.
% Cryptography and Distributed System solve this problem.

\subsubsection{Agents in Game Theory and Security}
Both Game Theory and Security deal with the interaction of agents, but they differ noticeably in how they model these agents. 
Security deals with \textit{adversaries}, agents that aim to circumvent or break a security property of the system. 
The value that an adversary attaches to their success is not usually given, as security should ideally be robust to any adversary, although they may have computational limitations.
Game Theory, on the other hand, deals with \textit{rational} (sometimes also called \textit{selfish}) agents that assign a value to their goals, and would rather optimize their payoff than achieve an arbitrary goal, but they typically do not have restrictions (e.g.,  computational) like a security adversary would.

In practice, this translates to different assumptions being used when formally modelling a game or the security of a system, making it difficult to prove statements involving security and game theoretic properties.
As both deal with different types of agents, a proof will involve complexity from both sides and quickly become hard to manage, leading to a disjoint treatment of both aspects in works that attempt to cover both.
There are nonetheless inherent connections as security often uses game based proofs, although an adversary wins if they have a high enough probability of succeeding in their attack rather than if their utility is high enough. 
But if we assume that the adversary has a high payoff associated with the success of their attack, then we start to recover game theoretic intuition.
\\ \indent For proofs based on the simulation (ideal/real world) paradigm, some connections are also evident.
The idea behind the simulation paradigm first originated in the context of secure computation.
Goldreich et al.~\cite{Goldreich:1987:PAM:28395.28420} introduced the idea of bypassing the need for a trusted third party (i.e., a mediator) in games of incomplete information, by replacing it with a protocol that effectively simulates it such that any information known by players at any step of game is the same as they would have known in an execution of the game involving the trusted third party.
This was refined by Micali and Rogaway~\cite{micali1991secure} in terms of ideal and secure function evaluation.
The ideal function evaluation corresponds to the evaluation of the function with a trusted third party that receives the private inputs of the parties and evaluates the function before returning the result to each party, achieving the best possible result.
% Intuitively, this achieves the best possible result.
The secure function evaluation involves the parties trying to replace the trusted third party with a protocol, which is considered secure if the parties cannot distinguish between the ideal and secure function evaluations, meaning that an adversary is not able to gain anything significant.
%This means that the protocol performs close enough to the standard of the ideal function evaluation, and an adversary would not have anything to gain.
Simulation has become a powerful tool for cryptographers~\cite{EPRINT:Lindell16} and Canetti's Universal Composability Model~\cite{FOCS:Canetti01} expands on these ideas to provide a framework for secure composability of protocols.

\subsubsection{Decentralization, incentives and security}
Because the security of most decentralized systems,
like cryptocurrencies, is linked not only to the security of the protocols, but also to 
%achieving an honest majority of participants, 
having a majority of participants following the rules,
decentralization and incentives have to be considered.

The ideal system does not depend in any way on any single party, which requires it to be decentralized.
Troncoso et al.~\cite{SystematizingDecentralizationandPrivacyLessonsfrom15YearsofResearchandDeployments} give an overview of decentralized systems, defining a decentralized system as ``a distributed system in which multiple authorities control different components and no single authority is fully trusted by all others''.
This highlights the fact that every component of the system should be decentralized, and in particular a single authority distributing its own system (or component) is not decentralized.
This can be hard to achieve in practice, and the level of decentralization of a system should always be looked at critically.
A decentralized system where all important parties are independent but under the jurisdiction of a single government may not truly be decentralized. 
All these independent parties may also depend on a very few hardware manufacturers (or other service providers).

Incentives are key to achieving an honest majority.
Azouvi et al.~\cite{10.1007/978-3-030-03251-7_15} give an overview of the role incentives play in security protocols, including cryptocurrencies, highlighting the fact that achieving guarantees of equilibria on paper may not be meaningful in practice when the wrong assumptions and models are used.
%, leading to unrealistic results.
What does security mean in this context? Clearly, protocols that are cryptographically secure, and that achieve safety (i.e., the guarantee that nothing bad will happend) and liveness (i.e., the guarantee that somethind good will happen) are needed, otherwise nothing else would work. 
But if the security of the system also depends on achieving a high enough degree of decentralization, more is required.
%In particular, decentralization relates to the users of the system rather than a protocol in the system. 
In particular, decentralization relates to the participants and their behavior rather than
solely the protocol.
Any decentralized protocol can always be ran in a centralized manner, so it is not enough to design a system that can be used in a decentralized manner. Rather, the requirement is to design a system that is advantageous to use in decentralized manner. 

Doing this naturally requires a better understanding of why users would want to be decentralized rather than try to gain more individual control of the system for themselves, so security is no longer just about the protocol itself, but also about how it can be used and how it is used.

%\subsubsection*{Bitcoin and cryptocurrencies}
% Another system that solves the Byzantine agreement problem
% is Bitcoin. In Bitcoin, the assumptions are very different
% from traditional distributed systems~\cite{bano2017consensus},
% as the setting is completely open and permissionless.
% Thus, it inspired a whole new direction in this field.
Bitcoin represents an important innovation from classical consensus protocols as it is fully open and decentralized.
In the Bitcoin consensus protocol (sometimes called Nakamoto consensus)
participants can join and leave as they wish, and Sybils are handled through the use of \textit{Proof-of-Work} (PoW).

The data structure that keeps track of the state of the system in Bitcoin is a chain of chronologically ordered blocks i.e.,  the blockchain, with each block containing a list of transactions.
To win the right to append a block (and win the block reward), participants compete to solve a computational puzzle i.e.,  a PoW.
They include in their block the solution to that puzzle, the PoW, such that other players can verify its correctness.
This block then initiate a new puzzle to be solved. This process of creating new blocks is called \textit{mining} and participants in this protocol are called \emph{miners}.

%If two participants find a solution at roughly the same time, two blocks will be created at the same height in the chain, creating a \textit{fork}, which is problematic as the two blocks can contain conflicting transactions.
%To resolve this, each participants will choose one of the two blocks and start mining on top of it by solving the associated PoW.
%Whenever a chain becomes longer than the other, participants will abandon the shortest one.
%As it is unlikely that two chains will keep the same length for a long time, participants
%then reach consensus by following the longest chain rule.

The security of Bitcoin relies on a majority of the mining power (i.e.,  hashing power) in the network following the protocol, whether because they are honest or simply rational.
Taking control of half of the computational power of Bitcoin for only an hour
has a considerable cost (around 670k USD~\cite{crypto51} as of May 2019), although it is with reach of potential adversaries.
This cost depends on the hash rate of the network (i.e.,  the cost of mining) and the price of Bitcoin in USD as once mining is no longer profitable for some miners they are likely to stop mining, reducing the hashing power required to control a majority of the network.

This is one of the reason why Bitcoin's security is so tightly linked to incentives, as when mining is no longer worthwhile the security of the network decreases.
Participation in the network is also rewarded by financial
gain (through block rewards and transaction fees).
The more participants there are, the harder it is to attack the network since the cost for mounting a 51\% attack (where an adversary takes control of more than half of the computational power) increases.
These financial motivations are thus also paramount. 

Since Bitcoin's deployment, many alternative cryptocurrencies that similarly rely on a blockchain have emerged.
The most popular of these is Ethereum~\cite{ethereum}, which differs from Bitcoin in that it provides a more complex scripting
language meaning that rather than processing simple transactions, nodes in the
system execute a script that allows users to perform multitude of functionalities
(so-called \textit{smart contracts}). 

\section{Cryptography and Game Theory}
\label{sec:cryptoandgt}
% In MD, the designer creates a game where
% rational players act in a way that
% gives an overall optimal outcome.
% Players can lie to the center but they still
% follow the rule in the sense that they choose
% among all the possible actions that are offered to them by the center.
% % however
% % they could lie (for example give a false value for their type).
% % but they cannot,
% % for example, learn about other people type.
% A cryptographic protocol provides a solution to obtain similar guarantees
% without any trusted third party.
% For example Secure Computation essentially guarantees that
% whatever computation $n$ players can do with the help of
% a trusted party, they can also do by themselves.
% Mechanism Design uses incentives to make the players behave in the optimal way
% Cryptography uses mathematics to force the players to behave correctly.
% For this reason Cryptography and GT go well together
% and cryptographers have started given some attention to GT problems.

% There are different approaches that have been taken when combining
% Cryptography and Game Theory. One is to add incentives and utility functions
% to standard cryptographic security notion and the other is on the opposite
% to start from standard game theoretic notions and add computational
% considerations.
This section considers work at the intersection
of Game Theory and Cryptography.
Cryptography usually considers a worst-case adversary, but by relaxing this assumption, it is possible to design protocols that bypass impossibility results or achieve better efficiency than existing ones, while maintaining a realistic adversarial model.
% On the other hand, in game theory all the games require the existence of a trusted
% third party to enforce the action space and payoffs. This is an unrealistic or
% even undesirable property that Cryptography can solve.
%First, in Section~\ref{sec:cryptogt}, we introduce the subfield of
%rational cryptography that considers a rational adversary and incorporates game theoretic notions
%like utility functions to cryptographic schemes. %, in order to achieve this.
%We particularly focus on the Rational Protocol Design
%framework of Garay et al.~\cite{rationalcrypto}. The approach taken here is to
%combine the UC framework (introduced in Section~\ref{sec:prelim}), with
%some mechanism design (MD) notions.
%Next, in Section~\ref{sec:gtcrypto}, we look at another approach that consists
%in adapting game theoretic notions to consider computational aspects in games.

\subsection{Cryptography meets Game Theory: Rational Cryptography}
\label{sec:cryptogt}

Initiated by Dodis, Rabin and Halevi~\cite{rationalcrypto3}
rational cryptography is a subfield of cryptography that incorporates incentives
in cryptographic protocols.
In this context, new adversaries and their capabilities have to be defined, as well as how to account for incentives and how protocols can be proven secure for such adversaries.

% In this new setting, the questions that arise are how to define the corresponding new adversary
% and what are their capabilities and limits; how to take into account their incentives; how to incorporate that novel adversary in the proof model;
% to which solution concepts or equilibria should it map to?

First, we note that most of this
work~\cite{rationalcrypto4,kol2008games,kol2008cryptography,gordon2006rational,fuchsbauer2010efficient,gt-secure-computation,dodis2007cryptography} focuses on multi-party secret sharing or secure function evaluation.
Thus, no monetary incentive is usually considered.
As pointed out by Dodis and Rabin~\cite{dodis2007cryptography}, in a rational cryptographic context, the utilities
of the players are usually dependent on cryptographic considerations such as:
correctness (a player prefers to compute the function correctly),
exclusivity (a player prefers that other players do not learn the value of
the function correctly),
privacy (a player does not want to leak information to other players),
voyeurism (a player wants to learn as much as possible about the other parties).

In addition to the above, other interesting parameters can come to play in the adversary's utility function.
For example, Aumann and Lindell~\cite{aumann2007security} formalized the concept of \textit{covert adversaries}
that may deviate from the protocol but only if they are not caught doing so.
As they argue, there are many obvious situations where parties cannot afford the effect of being caught cheating.
Covert adversaries are somehow similar to adding a punishment to the utility function.
Rational players do not want to be caught cheating as the punishment decreases their utility.
% \mygap{This framework does not provide any modular composition theorem unlike the UC
% framework.}
In Aumann and Lindell's setting the protocol detects the cheating, but in practice we need to incentivize participants to do so. %, otherwise cheating can go undetected.
Some work considers adding adversarial behavior together with rational 
adversaries~\cite{rationalityandadversarial}, we consider this further in Section~\ref{sec:dsgt}.

In terms of equilibria, the solution concepts proposed in these works are often extensions of a Nash Equilibria (NE), introduced in Section~\ref{sec:prelim}.
For example, Halpern and Teague look for a NE that remains after other NE that are weakly dominated (i.e., at best only as good as others) are removed through iterated deletion, where all dominated strategies are removed at each step~\cite{rationalcrypto4}.
%\saraha{@alex could you quickly explain iterated deletion of weakly ...}
Asharov et al.~\cite{gt-secure-computation} adapt the simulation-based definition to capture game-theoretic notions of (for example) fairness, meaning that one party learns the output of a computation if and only if the other does as well.
%The approach they take is to add a utility function for each notion of security considered (e.g., for correctness the utility will be $1$ if the output is correct and $0$ otherwise).
%They then show an equivalence theorem that states, roughly, that following the protocol is a Nash equilibrium if the protocol correctly computes the multi-party function in the presence of fail-stop adversaries.
%We present their framework in the following section.
% \saraha{theorem: Let f be a deterministic two-party function, and let $\pi$ a two-party protocol. Then,
% $\pi$ is a Nash protocol with respect to u
% and D
% if and only if $\pi$ correctly computes f in the
% presence of fail-stop adversaries.}
As their notions are weaker than standard cryptographic definitions, they can be achieved in some settings where impossibility results usually hold in traditional cryptography.

%\paragraph*{Rational Protocol Design}
% Garay et al.~\cite{rationalcrypto} also consider
% adapting simulation-based definitions with utility functions but while
% taking a novel approach.
Following the work presented above, Garay et al. propose Rational Protocol Design (RPD)~\cite{rationalcrypto}.
%follows the line of research
%of rational cryptography. % , where each party in a multiparty computation
% is treated as a rational agents trying to maximize their expected utility.
In this setting, they define a game between the designer of the protocol $D$
and the attacker $A$.
%The number of parties $n$ is known to $A$ and $D$.
The game is parametrized by a multi-party functionality $\mathcal{F}$ and
consists of two sequential moves.
In the first step, $D$ sends to $A$ the description $\pi$ of the protocol
that honest parties are supposed to execute. In the second step,
$A$ chooses a polynomial-time interactive Turing machine (ITM) $\adv$ to attack the protocol.
%The corresponding game is noted $\mathcal{G}_{\mathcal{M}}$, where $\mathcal{M}$
%corresponds to the attack model which specifies the functionality, the action
%sets and utilities.
The game is zero-sum in the original paper, but was later adapted to be a non-zero sum game in the context of Bitcoin~\cite{butwhy}.

%\subsection*{Defining the utilities}
%The methodology used to define the attacker's utility consists of three steps.
%First, relaxing the functionality $\mathcal{F}$ to $\langle\mathcal{F}\rangle$ which explicitly allows for some security breaches.
%Second, defining the payoff of any ideal-world adversary as a function $v$ of the view of any ideal evaluation of the relaxed functionality.
%Third, assigning to each adversarial strategy for a given
%protocol the expected payoff achieved by the best simulator.
%The best simulator here is the simulator that successfully emulates $\adv$
%while achieving the minimum score (the idea being that the adversary should be rewarded only if it forces the simulator
%to provoke an event).

%the simulator provokes an event even if it could
% simulate all necessary messages without doing so. The adversary on the other
% hand should be rewarded only if it forces the simulator to provoke the event.
%In the case of a zero-sum game, the utility of the designer $D$ is then
%straightforwardly defined as the opposite of the adversary's.

In follow-up work~\cite{fairness-rdp}, notions
of fairness are also considered, and provide a mean of comparison between protocols i.e., which protocol is the fairer.
Informally, a protocol $\pi$ will be at least as fair as another protocol $\pi_0$
if
the utility of the best adversary $A$ attacking $\pi$ (i.e., the adversary which maximizes $u_A(\pi, A))$ is no
larger than the utility of the best adversary attacking $\pi_0$, except for some negligible quantity.

%\subsection{Solution Concepts and Security Definitions}
The solution concept introduced
within the RPD framework is $\epsilon-$subgame perfect equilibrium where the parties' utilities are
$\epsilon$ close to their best response utilities.
% More formally it is defined as follows:
% \begin{definition}[$\epsilon-$subgame perfect equilibrium]
% Let $\mathcal{G}_{\mathcal{M}}$ be an attack game. A staretgy profile $(A,\Pi)$
% is an $\epsilon-$subgame perfect equilibrium in $\mathcal{G}_{\mathcal{M}}$ if:
% (1) for any $\Pi'\in \trm{ITM}^n$, $u_D(\Pi',A(\Pi'))\le u_D(\Pi,A(\Pi))+\epsilon$,
% and (2) for any $A'\in \trm{ITM}$, $u_A(\Pi,A'(\Pi))\le u_A(\Pi,A(\Pi)) +\epsilon$.
% \end{definition}
When it comes to security, the RPD framework defines the
notion of attack-payoff security.
Informally, attack payoff security states that an adversary has no
incentive to deviate from the protocol.
%formally it is defined as follows:

Another concept, incentive compatibility, was introduced
in a follow-up work of RPD~\cite{butwhy}. Here, the definition is slightly different
than the definition usually given within Mechanism Design (MD) where participants
achieve the best outcome by revealing their true preferences.
Informally, incentive compatibility states that agents gain some utility
when participating in the protocol i.e., they choose to play instead of ``staying at home''.

% The security in the RPD is mostly captured through the simulation,
% the proof needs to show that one can UC-realize the
% relaxed ideal functionality. Note that here the ideal functionality is
% ``relaxed'' to allow for some potential attacks.
% These attacks are then thwarted with an incentive argument:
% no adversary has an incentive to exploit the vulnerabilities.

%To sum up, a protocol is secure in the RPD framework if it UC-emulates the relaxed functionality
%and an adversary has no incentive in exploiting the attack.
%% Showing that a protocol is secure in the RPD framework implies showing
%% the UC-emulation of the relaxed functionality and showing that
%%Proofs are very tedious in that framework. 
%In their work on Bitcoin~\cite{butwhy}, Badertscher et al. state that it is only necessary to consider the
%real world and not the ideal world in proofs. This makes us wonder whether this framework could be simplified more.

Apart from the recent work of Badertscher, et al.~\cite{butwhy}, rational cryptography does not
consider monetary payment.

One important drawback of RPD is that it does not consider the presence of irrational adversaries despite the fact that in security, we do not always know the motivation of an attacker.
RPD uses a relaxed functionality to allow for some defined attacks but this may not cover all attacks, leaving the door open to potential attacks.
%Would there be a way to take a similar approach without having to relax the functionality in order to account for every attack?
%The UC model is meant to account for everything that could happen in the environment (i.e., the universe of the protocol) but it is a computational model, when they add incentives into their consideration, 
The UC model does not automatically start accounting for all possible incentives - this is a clear flaw as we know that only arbitrarily considering incentives leads to failures (e.g., failure of considering outside incentives, or in general "soft" incentives like political and other external incentives~\cite{10.1007/978-3-030-03251-7_15}). 

\subsection{Game Theory meets Cryptography: computational games}
\label{sec:gtcrypto}
%The classic MD literature largely ignores computational considerations.
%The challenge is to make mechanisms computationally feasible without sacrificing useful
%game-theoretic properties, such as efficiency and strategy-proofness.

%\subsection{Computational games}
Rather than starting from a cryptographic setting and incorporating game theoretic notions, as presented above, one can also start from a game theoretic setting and from there move towards cryptographic notions by considering the computational aspects of games.
This approach is taken in a body of work by Halpern and Pass that considers \textit{Bayesian machine games} first introduced in a preprint~\cite{2008arXiv0809.0024H} that has later appeared in different forms~\cite{Pass:2010:GTC:1807406.1807495,Halpern:2011:ARA:1998549.1998551,halpern2015algorithmic}, primarily venues focused on Economics rather than Security.

A Bayesian machine game (BMG) is defined very similarly to a standard Bayesian game (introduced in Section~\ref{sec:prelim}), it only differs in that it considers the complexity (in computation, storage cost, time or otherwise) of actions in the game.
This is done by having players pick machines (e.g., a TM or ITM) that will execute their actions and defining a complexity function for that machine, which the utility takes into account.
%Correspondingly, player types are restricted to $\{0,1\}^*$.
%While any model of computation can in principle be chosen to represent the machines, the usual choice is that of Turing Machines (TM), or Interactive Turing Machines (ITM) if it is necessary for the machines to receive messages, as is practice in Cryptography.

% \begin{definition}[Bayesian Machine Game~\cite{2008arXiv0809.0024H}]\label{def:bmg}
% % A Bayesian machine game G is described by a tuple $(N, \mathcal{M},\Theta,\Pr,\mathcal{C}_1,\dots,\mathcal{C}_m,u_1,\dots,u_2)$ where:
% % \begin{itemize}
% % 	\item $N$ is the set of players, $\mathcal{M}$ is the set of possible machines
% % 	\item $\Theta\subseteq(\{0,1\}^*)^{m+1}$ is the set of type profiles where the $(m+1)$st element in the profile corresponds to nature's type
% % 	\item $\Pr$ is a distribution on $\Theta$
% % 	\item $\mathcal{C}_i$ is a complexity function
% % 	\item $u_i:T\times(\{0,1\}^*)^m\times\mathbb{N}\rightarrow\mathbb{R}$ is player i's utility function.
% % \end{itemize}
% % \end{definition}

A Nash equilibrium for a BMG is expressed in the usual way, but it now takes into account the machine profile rather than a strategy profile.
% \begin{definition}[Nash equilibrium in machine games~\cite{2008arXiv0809.0024H}]
% Given a Bayesian machine game G, a machine profile $\vec{M}$, and $\epsilon\geq0$, $M_i$ is an $\epsilon$-best response to $\vec{M}_{-i}$ (the tuple consisting of all machines in $\vec{M}$ other than $M_i$) if, for every $M^{'}_{i}\in\mathcal{M}$,
% \begin{equation}
% U^G_i[(M_{i},\vec{M}_{-i})]\geq U^G_i[(M^{'}_{i},\vec{M}_{-i})]-\epsilon.
% \end{equation}
% $\vec{M}$ is an $\epsilon$-Nash equilibrium of G if, for all players i, $M_i$ is an $\epsilon$-best response to $\vec{M}_{-i}$. A Nash equilibrium is a $0$-Nash equilibrium.
% \end{definition}
There is, however, an important distinction to make between a standard Nash equilibrium and a Nash equilibrium in machine games, which is that the latter may not always exist.
The necessary conditions for the existence of a Nash equilibrium in a machine game are given by Halpern and Pass~\cite{2008arXiv0809.0024H} to be a finite type space, bounded machines and a computable game.
% \begin{theorem}[Existence of Nash equilibrium in a machine game~\cite{2008arXiv0809.0024H}]
% If $\Theta$ is a finite type space, $B$ is a bounding function, $\mathcal{M}$ is a set of $B$-bounded machines and $G=(N, \mathcal{M},\Theta,\Pr,\vec{\mathcal{C}},\vec{a})$ is a computable, computationally cheap Bayesian machine game, then there exists a Nash equilibrium.
% \end{theorem}
A follow up paper by Halpern et al.~\cite{2015arXiv150701501H} discusses the general question of the existence of a Nash equilibrium for resource bounded players.

So far, the discussion of computational games has not yet touched on security related issues, but Halpern and Pass prove an equivalence theorem that relates the idea of universal implementation in a BMG to the standard notion of secure computation in Cryptography~\cite{Ben-Or:1988:CTN:62212.62213,Goldreich:1987:PAM:28395.28420}.
Intuitively, this goes back to the work of Goldreich, Micali and Widgerson~\cite{Goldreich:1987:PAM:28395.28420} that first expressed (to the best of our knowledge) the idea of secure computation as the replacement of a mediator in a game that preserves an equilibrium.

A universal implementation corresponds to the idea that a BMG implements a mediator if whenever a set of players want to truthfully provide their input to the mediator, they also want to run their machine using the same input, preserving the equilibrium and action distribution.
% More formally it is defined as follows.
% \begin{definition}[Universal Implementation~\cite{2008arXiv0809.0024H}]
% Suppose that $\mathcal{G}$ is a set of n-player canonical games, $\mathcal{Z}$
% is a subsets of $N$, $\mathcal{F}$ and $\mathcal{F}'$ are mediators,
% $M_1,\cdots,M_n$ are interactive machines, $p:\mathbb{N}\times\mathbb{N}\rightarrow\mathbb{N}$
% and $\epsilon:\mathbb{N}\rightarrow\mathbb{R}$.
% $(M,\mathcal{F}'$ is a $(\mathcal{G},\mathcal{Z},p)$-universal implementation
% of $\mathcal{F}$ with error $\epsilon$ if, for all $n$, all games $G\in\mathcal{G}$
% with input length n and all $\mathcal{Z}'\subseteq\mathcal{Z}$
% if $\vec{\Lambda}^\mathcal{F}$ is a $p(n,\cdot)$-robust $\mathcal{Z}'$-safe $\epsilon$-NE
% in the mediated machine game $(G,\mathcal{F})$ then
% \begin{enumerate}
% \item (Preserving equilibrium) $\vec{M}$ is a $\mathcal{Z}'$-safe $\epsilon$-NE
% in the mediated machine game $(G,\mathcal{F}')$
% \item (Preserving Action Distributions) For each type profile $\vec{t}$,
% the action profile induced by $\vec{\Lambda}^\mathcal{F}$ in $(G,\mathcal{F})$
% is identically distributed to then action profile induced by M in $(G,\mathcal{F}')$.
% \end{enumerate}
% \end{definition}
There are then multiple equivalence theorems of different strength (up to the information theoretic case), that relate flavors of secure computation to flavors of implementation. The relation is important, as it not only implies that secure computation leads to a form of game theoretic implementation, but also the reverse. This opens up the option that the guarantees of (some flavor of) secure computation could be achieved by considering the Game Theory of a problem, although it is not clear whether this process would be more efficient.
BMG have natural applications to known security problems. For example, dealing with covert adversaries as described by Aumann and Lindell~\cite{aumann2007security} (introduced above) can be done by introducing a (two player, for example) mediated game where the honest strategy is to report your input to 
the mediator and output its reply (with utility $\frac{1}{2}$), and the string \textit{punish} can be output by a player to ensure the other receives payoff $0$. Then any secure computation with respect to covert adversaries with deterrent (probability of getting caught cheating) $\frac{1}{2}$ is an implementation of the mediator as the expected utility of a cheating player will be $\frac{1}{2}\cdot1+\frac{1}{2}\cdot0=\frac{1}{2}$, which is the same as that of the honest strategy.
\section{Distributed Systems and Game Theory}
\label{sec:dsgt}

% In the case of large decentralized systems, it can be hard
% to force players to behave in a given way so relying on well
% aligned incentives is necessary for the system to work.
In this section, we present the work that is at the intersection
of Game Theory and Distributed Systems, looking at concrete problems that
have been well studied.  
For each case we will illustrate important
concepts and techniques used.
\subsection{Algorithmic Mechanism Design}
Algorithmic mechanism design (AMD) is concerned with designing games such that self-interested players achieve the game designer's goals, in the same way that distributed systems designers aim, for example, to achieve agreement in the presence of Byzantine players.
AMD was first introduced by Nisan and Ronen~\cite{nisan2001algorithmic} who proposed that an algorithm designer should ensure that the interests of participants in a distributed setting are best served by behaving correctly i.e., the algorithm  designer should aim for incentive compatibility. 
% A key result was to show that the Vickrey-Groves-Clarke mechanism~\cite{nisan2001algorithmic}, a standard mechanism to implement the efficient outcome of a problem, could be used to solve design problems like the shortest path problem and task scheduling, which are relevant to issues like routing and load balancing.
The framework of Nisan and Ronen is defined for a centralized computation, but it has been extended to distributed algorithmic mechanism design (DAMD) following work by Feigenbaum et al. on cost sharing algorithms for multicast transactions~\cite{feigenbaum2001sharing}. 
This lead to further developments and applications of DAMD to interdomain routing, web caching, peer-to-peer file sharing, application layer overlay networks and distributed task allocation, which are summarized in a review by Feigenbaum and Shenker~\cite{feigenbaum2002distributed}.
% In this section we will illustrate what could go wrong in Distributed
% Systems where agents are rational with some concrete examples that
% have been well studied. 

\subsection{Public goods, free riding and hidden actions}
Public goods, which are produced at a cost but available to use for free, naturally occur in distributed systems.
In a public goods game, players choose to contribute a certain amount, with all contributions being combined and distributed among all players.
Naturally, this can lead to players rationally deciding to contribute less to maximize their utility.
They may even contribute nothing, which is generally referred to as \textit{free riding}.

Varian first considered modeling the reliability of a system as a public good~\cite{varian2004system}.
The reliability can either depend on the total effort (sum of the efforts exerted by the individuals), on
the weakest link (minimum effort) or on the best shot (maximum effort).
E.g., if there is a wall defending a city, its reliability can depend on the sum of all the work
provided by the builders (total effort), on the lowest height (weakest link) or if we consider several walls, on the highest one (best shot).
In the case of total effort  the NE corresponds to all players free riding on the player with highest benefit-cost ratio. Moreover, the effort exerted in the NE is always lower than the social optimum i.e., the best outcome across all players.

Peer-to-peer file sharing also provides an interesting case of a networked system that has faced free riding~\cite{feldman2005overcoming}.
% A notable example is that of peer-to-peer file sharing systems.
% A major problem when designing incentives for such systems is
% that of free-riding i.e., players consuming resources from the network without contributing to it~\cite{feldman2005overcoming}.
As explained by Babaioff et al.~\cite{babaioff2007incentives}, solutions to this problem could be based on a reputation system, barter or currency.
%In practice, these solutions are not always implemented in the system itself but rather added in an ad hoc way by users.
Another approach, that would not need to keep any long term state information is to replace \textit{indirect reciprocity} with 
\textit{direct reciprocity}.
For example, a file in BitTorrent is partitioned into smaller chunks, requiring
repeat interactions among peers and enforcing more collaboration between them~\cite{cohen2003incentives}.
In practice, however, this has been shown to not be very effective as it is not robust to strategic agents~\cite{piatek2007incentives} and induces free riding~\cite{jun2005incentives}.
% In practice, BitTorrent was designed with tit for that reciprocity in mind, but this has been shown not to work in theory and in practice~\cite{piatek2007incentives,jun2005incentives}.
% \saraha{sentence to summarize the problem with bittorrent incentives?}
% \alex{@Sarah: add something here, idk if it fits better in the paragraph above or below (or on its own)}
There is also the problem of dealing with newcomers, as an adversary can create new identities in order to abuse the system. Analyzing the incentives at play, Feldman et al.~\cite{feldman2006free} suggest that penalizing all newcomers may be an effective way of dealing with the problem, as it is not possible to penalize only users abusing the system.

In addition to free-riding, there are many other parameters that a selfish player could abuse in a P2P file sharing system e.g., when to join or leave,
who to connect to, untruthful sharing information, and so on.
This is the problem of \textit{hidden actions} i.e., how peers selfishly behave when their actions
are hidden from the rest of the network.
In order to analyze the degradation due to hidden actions, Babaioff et al.~\cite{babaioff2007incentives} apply the principal-agent framework,
due to the similarity of the hidden action problem with that of \textit{moral hazard}.
This framework is used in economics when one entity, the \textit{principal}, employs a set of $n$ agents to take actions on its behalf.
% (see Chapter 23 of~\cite{nisan2007algorithmic}).
%The principal will pay each agent to reward them for their effort, based on each observable action.
In order to capture the efficiency of a system in that framework, they define the \emph{Price of Unaccountability} of a technology as the worst ratio between the principal's utility in the observable-actions case and the hidden actions case.
%and the design of optimal contracts to induce efforts by the agents.
Dealing with observable and hidden actions relates to the transparency of the system, which can be approached from a cryptographic point of view
to ensure that agents all see the same set of actions~\cite{Chase:2016:TOA:2976749.2978404}.
% \alex{Inishgt: the observable actions is related to transparency, which is a cryptographic property~\cite{Chase:2016:TOA:2976749.2978404}}
%\mytechnique{The principal-agent Model}
 Another solution, Karma~\cite{Vishnumurthy2003karma},
proposes a system for peer-to-peer resource sharing that avoids free riding,
based on a combination of reputation system and consensus protocols. 
%It can been seen as a precursor of Bitcoin as it considers both a version of PoW and the idea
%of rewarding peers in the system for their effort.

%\subsection{Selfish Routing}
Another important problem in Distributed Systems where rationality can cause problems is routing~\cite{selfish-routing}.
The problem is to find a path that minimizes the latency between a source and a target.
One of the difficulties in doing so is that in a decentralized communication network it is not always possible to impose some routing strategy to nodes in order to, for example, regulate the load on a route.
As highlighted in our background on Game Theory in Section~\ref{sec:prelim}, nodes usually act according to their own interests, which can be orthogonal to the overall optimal equilibrium. 
A game theoretic measure used by Roughgarden and Tardos in the context of
routing is the \emph{Price of Anarchy}~\cite{selfish-routing}, which quantifies how much a system degrades due to selfish behavior.
More formally, assuming we have a measure of the efficiency of each outcome,
the Price of Anarchy is the ratio between the equilibrium and the optimal outcome.
 Inspired by this measure, Grossklags et al.~\cite{grossklags2010price} introduce the \emph{Price of Uncertainty}, which measures the cost of incomplete information compared to that of complete information.
An important observation is that assuming fixed possible losses, which is reasonable in the case of mining where one can at most lose the fixed cost hardware (and electricity) or stake, the more players are in the network, the less information matters. This also ties in to the value of information i.e., the possible change in utility from gaining information, which is defined for a computational setting by Halpern and Pass~\cite{halpern2010don}.

% It measures the degradation of a system of complete information compared to partial informatio,nand 
% is similarly defined as the ratio of some efficiency measure in these two cases.
% \saraha{add more about price of uncertainty}
% Takes into account the incentives of the agents.
% Follows from the work on algorithmic mechanism design,
% for a mechanism to be feasible in practive it has the outcome and payments function
% have to be tractable.
% One of the main motivation for DALMD is internet as it is decentralized.
% Decentralization also allows to do without a center.

% \begin{definition}
% A distributed mechanism $d^M$ is a 3-tuple $d^M=(\Sigma,g,s^M)$, where $\Sigma=(\Sigma_1,\cdots,\Sigma_n)$
% is the feasible strategy space of the nodes, $g:\Sigma\rightarrow O$ is the outcome function computed by the mechanism and $s^M=(s_1^M,\cdots,s_n^M)\in\Sigma$  is the prescribed strategy.
% \end{definition}

% \subsection{Solution Concepts}
% similar to the GT ones?
% \begin{definition}
% A strategy profile $s^*\in\Sigma$ is an ex-post Nash equilibrium of a distributed mechanism
% $d^M=(\sigma,g,s^M)$ if
% \[u_i(g(s_1^*(\theta_1),\cdots,s_n^*(\theta_n)),\theta_i)\ge u_i(g(s_1^*(\theta_1),\cdots,s_i^{'*}(\theta_i),\cdots,s_n^*(\theta_n))
% \]
% for every node $i$, for every possible strategy $s'_i\in\Sigma_i$, $\forall \theta_i$ and for all
% possible private types of the other node $\theta_j$
% \end{definition}

\subsection{Consensus}

% We start by giving a general overview of the work
% done in this field before
% introducing some relevant associated solution concepts and definitions~\ref{sec:dsgt-solution}.
% \subsection{Overview}
% \label{sec:dsgt-overview}
\subsubsection{Fault tolerance with rational players}
We now look at the example of consensus.
The approach here is
%, unlike the two previous examples where rationality was a problem to deal with,
to use incentives to bypass impossibility results on Byzantine Agreement
or improve on existing constructions.
% Different approaches have been considered.
In order to apply GT to DS, additional adjustments have to be made.
For example, traditional GT considers deviation from only one agent (as in a NE) while in practice agents form coalition.
In addition, in a DS it is important to consider multiple types of failures (e.g., processors
may crash) that are not considered in GT.
% Abraham et al. also extend existing results to consider players that behave in
% irrational or unexpected ways.
% This solution concepts is also applied in~\cite{leader-election-gt}

In order to account for both of these requirements,
%When it comes to the type of players
the BAR model defined by Aiyer et al.~\cite{barmodel} introduces three different types of players:
Byzantine, altruistic (players that simply follow the rules) and rational players.
In this case, the expected utility of a rational player is usually defined by
considering the worst configuration of Byzantine players and the worst set of strategies
that those Byzantine players could take, assuming all other non-Byzantine players obey the specified strategy profile. 
The goal of the BAR model is to provide guarantees similar to those of Byzantine fault tolerance to all rational and altruistic nodes, as opposed to all correct nodes.
Two classes of protocols meet this goal, Incentive-Compatible Byzantine Fault Tolerant (IC-BFT) protocols and Byzantine Altruistic Rational Tolerant (BART) protocols.
IC-BFT protocols, which are a subset of BART protocols, ensure that the protocol satisfies security and is the optimal one for rational nodes, while a BART protocol simply ensures security properties.
% An IC-BFT protocol thus must define the optimal strategy for a
% rational node. In a BART protocol a rational node may exploit local
% optimizations not specified in the protocol without endangering the
% global guarantees. 
% Note that IC-BFT protocols are a subset of the
% BART protocols.

Groce et al.~\cite{barational} introduce similar notions, perfect and statistical security,
which state that in the presence of a rational adversary, the protocol still
satisfies the security properties (e.g., consistency and correctness for consensus).
% For example, starting from the observation that traditional systems consider the worst case adversary,
% Groce et al.~\cite{barational}
% use a rational adversary that act in utility-maximizing way instead.
They show feasibility results of information-theoretic (both perfect and statistical) Byzantine Agreement,
assuming a rational adversary and complete or partial knowledge of the adversary preferences. 
Their protocols are also more efficient than traditional Byzantine Agreement protocols.

% The Algorithmic Game Theory book by Nisan et al.~\cite{nisan2007algorithmic} also features a more recent survey by Feigenbaum et al.~\cite{feigenbaum2007distributed}. 
% Going from mechanism design to AMD and DAMD requires reconsidering the assumptions about participants in a system (as outlined in Section~\ref{sec:prelim}) and computational restrictions related to the desired mechanism and its implementation. 
%With regards to participants, they are split into obedient, faulty, strategic and adversarial nodes of the network.
In the DAMD setting~\cite{feigenbaum2007distributed}, participants are split into obedient, faulty, strategic and adversarial nodes of the network.
%Obedient nodes are correctly functioning machines that have no strategic goals and thus simply perform what they are programmed to do.
%Faulty nodes are incorrectly functioning nodes that also do not have strategic goals but suffer from bugs or misconfiguration.
%Strategic nodes are selfish agents that aim to maximize their utility.
%Adversarial nodes are adversaries in the security sense of the word, ranging from honest but curious to Byzantine in their strategic goals.
The split follows the same lines as that of the BAR model, but separates the adversarial nodes from those that are faulty with no strategic goal.
Computational restrictions here are expressed with regards to the solution concepts rather than the agents.
This ties into topics in computational Game Theory, as a solution to a DAMD problem requires not only that incentive compatibility is achieved, but also that the solution be computationally tractable, which is not always the case.
(The tractability of computing Nash equilibria, or approximations, is out of the scope of this paper.)
The takeaway is that many solutions on paper are not straightforwardly obtained in an algorithmic setting, whether centralized or decentralized, and even approximations may not be enough.

\subsubsection{Robustness}
When it comes to adapting a NE to consider coalitions and irrational players
Abraham et al.~\cite{multiparty-computation-game-theory} extend the work of Halpern and Teague~\cite{Halpern:2004:RSS:1007352.1007447} to consider multiple players.
They introduce the concept of robustness that encompasses two notions, resilience and immunity.
Resilience captures the fact that a coalition of players has no incentive to deviate from
the protocol, and is similar to the concept of collusion-proof NE~\cite{beyond}.
Immunity captures the fact that even if some irrational players are present in the system, the utilities of the other players are not affected.
% i.e., their utility does not decrease.
An equilibrium that is both resilient to coalitions of up to $k$ players, and immune to up to $t$ irrational players is then said to be $(k,t)$-robust
% Robustness is defined more formally in the glossary.
% \alex{Should be a bit more precise here, since k,t robustness doesn't even appear but is mentioned in the next paragraph}

Robustness is a very strong property, but it is hard to achieve in practice.
Clement et al.~\cite{theoryofbargames} show that no protocol is $(k,t)$-robust if any node may crash and communication
is necessary and costly. When designing cryptocurrencies, however,
it is not unusual to consider that communication is free.

As discussed in Section~\ref{sec:cryptogt} with covert adversaries, it can be helpful to add a form of punishment to enforce correct behavior by rational players. 
%, i.e., if players are caught misbaheving their payoff will decrease.
Halpern et al.~\cite{multiparty-computation-game-theory} define a 
$(k,t)-$punishment strategy such that for any coalition of at most $k$ players and up to $t$ irrational players, as long as more than $t$ players use the
punishment strategy and the remaining players play the equilibrium strategy, then if up to $k$ players collude, they will be worse off than they would have been if the rational players had played the equilibrium strategy.
The idea is that by having more than $t$ players use the punishment strategy is enough to stop $k$ players colluding and deviating from the equilibrium strategy.

% everyone not in $T$ played $\sigma$. The idea is that
% the threat of having more than $t$ players use their component
% of $\rho$ is enough to stop players in C from deviating from $\sigma$.''
% See the glossary for a formal definition.

\subsubsection{Price of Malice} 
As systems realistically involve rational and irrational players, it is important to consider how rational players react to the
presence of irrational players.
Moscibroda et al.~\cite{selfishmeetsevil} do this by considering a system with only rational and Byzantine players. 
They differentiate between an \emph{oblivious} and \emph{non oblivious model} i.e., whether selfish players know the existence of Byzantine players or not.
They define a \textit{Byzantine Nash equilibria} that extends NE in the case where irrational players are present.
In a Byzantine Nash equilibria no selfish player can reduce their perceived expected cost, which depends on their information, by changing their strategy, given that the strategies of all other players are fixed.

In GT and MD, a concept very often discussed is
the Price of Anarchy~\cite{tim-survey}, which was introduced in the case
of selfish routing.
Moscibroda et al.~\cite{selfishmeetsevil} extend this to their setting, by defining the \textit{Byzantine Price of Anarchy} that quantifies how much an optimal system degrades due to selfish behavior, when malicious players are introduced.
More formally, it is the ratio between the worst %and best case 
social cost of a Byzantine Nash equilibrium and the minimal social cost, where the social cost of a strategy profile is the sum of all individual costs i.e., the optimality of each outcome.

The price of Malice is used to see how a system of purely selfish players degrades in the presence of malicious irrational players.
More formally it is the ratio between the worst Byzantine Nash Equilibrium with malicious players and the Price of Anarchy in a purely selfish system. 

Moscibroda et al.~\cite{selfishmeetsevil} also introduced the idea that Byzantine players can
improve the overall system, which they called the \textit{fear factor}.
The intuition is that the rational players will adapt their strategies by fear of the actions of irrational players, rendering the overall system better.
The example they introduce where this can be observed is virus inoculation.
Based on the assumption that some players are irrational and will not get vaccinated, rational players will be incentivized to get vaccinated.
In the case where everyone is rational, there is no equilibrium since as long as enough people get vaccinated, the rest of the population is safe.
Thus irrational players here can make the overall system better.

% \subsection{Discussion}

% Most of the new definitions considered consist in having the same security guarantees
% as before even in the case where some players are selfish (BART, perfect/statistical security).
% Robustness is a general extension of NE in the DS context.
% Price of malice is an interesting new measure to quantify how selfish 
% and Byzantine behaviours can degrade the system away from its optimal state.
% The fear factor is a very interesting concept as well,
% as it captures how players react when they think
% irrational players are here. This is tightly related to the notion of knowledge
% in GT.

%!TEX root = main.tex
\section{Blockchains}\label{sec:blockchains}
We now consider blockchain based cryptocurrencies, which are an important example of systems involving aspects of both traditional security and game theoretic aspects. 
%Incentives here are explicit as they are monetary and coupled with security.
%% The security of Bitcoin relies on the proof-of-work
%Participants are incentivized to participate in the network (i.e., become a miner)
%due to the financial reward associated with it. Moreover, participants
%are incentivized to ``follow the rules'' due to the cost of creating blocks and risk of financial loss with deviating from the rules.
%
%Despite this, and probably due to their recent apparition the community has not agreed on a formal model that fully incorporates incentives and security, although some work is being done in this direction~\cite{smartpool,solidus,fantomette} to various degrees of success.
%Existing work has highlighted failures in incentive models e.g., selfish mining~\cite{selfishmining}, the verifier's dilemma~\cite{verifierdilemma} and the miner's dilemma~\cite{minersdilemma}.
In this section we review the work that has been done by the security and distributed systems communities on blockchains that consider game-theoretic notions.

We start by reviewing the work that give some game-theoretic analysis
of Bitcoin. We then illustrate how these analysis fell short with
attacks that have been found on Bitcoin's incentives.
We then gove an overview of the work that has been done on blockchain consensus protocols that considers the question of incentives and try to thwart these attacks.
As we argued in Section~\ref{sec:background}, having a system that is secure
with some bounded number of Byzantine faults is not enough to have a decentralized system as decentralization cannot be assumed. 
Rather, incentives should be designed to ensure enough participation and prevent coalitions.
We therefore also discuss work focusing on incentivizing decentralization.
We then review the work on payment channels before finally presenting some
notions of fairness with respect to blockchains' reward systems.

Along the way, we also highlight new concepts of interest that have been introduced in this field as well as how they relate to what we have previously discussed in this paper.

%Well many incentive related papers present attacks i.e., deviations to the protocol that rational agents may follow~\cite{minersdilemma,selfishmining,optimalselfishmining,verifierdilemma}), with fewer papers focusing on proving security while incorporating incentives~\cite{butwhy}. 
%This general approach of proposing an attack and a patch to said attack is similar to the one taken in Cryptography before provable security existed.

% We divide this section as follows: we first present attacks on incentives
% and general models that try and prove security with incentives for Bitcoin.
% We then present new blockchain protocols design that consider incentives and their
% models. Next, we present the work that has been done focusing on incentivizing
% decentralization. Lastly we talk about protocols build on top of blockchains
% that also rely on economic argument for their security.

%Some papers consider NE extensions:

\subsection{Game theoretic analysis of Bitcoin}
Nakamoto's original Bitcoin paper~\cite{satoshi-bitcoin} provided only informal security arguments but several papers have since formally argued the security of Bitcoin in different models~\cite{pass2017analysis,Kiffer:2018:ccs,bitcoin-backbone}, usually based in the simulation setting presented in Section~\ref{sec:prelim}, but without a consideration of incentives.
%All of these models are based on the simulation paradigm used in Cryptography (see Section~\ref{sec:prelim}).

In early work in this area Kroll et al.~\cite{economicsbitcoin} show that there is
a NE in which all players behave consistently with Bitcoin's reference implementation, along with infinitely many equilibria in which they behave otherwise e.g., where they all agree to change a rule. 
Attacks like selfish mining~\cite{selfishmining,optimalselfishmining,nayak2016stubborn} put this into question, showing that their model did not encompass behavior that could realistically occur.
More recently, Fiat et al.~\cite{fiat2019energy} showed that the only possible pure equilibria in Bitcoin's mining are
very chaotic (miners quitting and starting again periodically) or non-existent, depending on the configuration of players..

More recently Garay et al.~\cite{butwhy} proved the security of Bitcoin in the RPD framework that was introduced in Section~\ref{sec:cryptogt}.
Their approach is based on the observation that Bitcoin works despite its flaws, and they prove that Bitcoin is secure by relying on the rationality of players rather than an honest majority.
This model inherits the flaws discussed in Section~\ref{sec:cryptogt} e.g., they do not consider fully malicious players.
Their model also does not encompass attacks on Bitcoin's incentive structure that we now describe.
%For example, miners are rational, they do not consider fully malicious players.
%Furthermore, this model still does not encompass the attacks found on Bitcoin's incentives that we describe now.

\subsection{Attacks on incentives}
With time, many attacks related to incentives in cryptocurrencies have been found, typically involving either external incentives (e.g., in bribery attacks) or the unintended use of a cryptocurrency's technical mechanisms. The effect of these attacks results in lowering the power required for a 51\% attack to less than 51\%.

Selfish mining~\cite{selfishmining} involves a rational miner increasing their expected utility by withholding their blocks instead of broadcasting them to the rest of network, giving them an advantage in solving the new proof-of-work and making the rest of the network waste computation by mining on a block that is not the top of the chain.

Inspired by techniques introduced by Gervais et al.~\cite{Gervais:2016:SPP:2976749.2978341}, Sapirshtein et al.~\cite{optimalselfishmining} use Markov Decision Processes (MDP) to find the optimal strategy when doing selfish mining. (MDP are used to help make decisions in a discrete state space where outcomes are partially random.) They show that with this strategy, an adversary could mount a 51\% attack with less than 25\% of the computational power.
This problem is further studied by Hou et al.~\cite{hou2019squirrl} using deep
reinforcement learning. They suggest that selfish mining becomes less effective when
performed by multiple adversaries.
In addition to witholding their own block, miners are neither incentivize
to propagate information (e.g. transactions or blocks) to the rest of network.

This is a problem that also exist in any P2P systems~\cite{emek2011mechanisms,li2009incentive} that researchers have also looked into solving, using techniques similar to those proposed in Distributed Systems~\cite{babaioff2012bitcoin,solidus,ersoy2018transaction}.

% In an MDP, a player moves through a
% discrete state space and tries to maximize reward
%actions are limited and random
%in situations where outcomes are partly random and partly under the control of a decision maker. 

Another issue is the verifier dilemma~\cite{verifierdilemma}, which shows that miners are not incentivized to verify the content of blocks, especially when this incurs an important computation on their end.

Mining gaps are another type of attack on incentives~\cite{instability-bitcoin,mining-gap} where the time between the creation of blocks increases because miners wait to include enough transactions (in order to get the transaction fees).
%Both papers use simulations to quantify the attacks, using techniques such as no-regret learning
%where miners update their strategy at every ``stage'' of the game in a way to do as close to the best strategy
%as possible, had it be known from the beginning or MDP.
%  we would like to select a sequence of strategies that does nearly as well as the best strategy, assuming
% we knew it from the beginning. 

Bribery attacks are another family of attacks that are often thought of as an example of the \textit{tragedy of the commons}, which describes a situation when individuals acting selfishly affect the common good~\cite{hardin1968tragedy}.
In our context, it captures the fact that miners have to balance their aim to maximize their profit with the risk of affecting the long term health of the cryptocurrency they mine, potentially reducing its price and their profit.
% This is related to the idea of the \textit{tragedy of the commons},  that expresses the fact that miners have to balance maximising their short term profit with the risk of affecting the long term health of the cryptocurrency they mine.
% \saraha{@alex je comprends pas bien la phrase precedente}
%\alex{I'd like to say something about how miners can also switch cryptocurrencies that require the same hardware but that isn't the right place}

Bonneau~\cite{FCW:Bonneau16a} first proposed that an adversary could mount a $51\%$ attack at a much reduced cost by renting the necessary hardware for the length of the attack rather than purchasing it.
More generally, a briber could pay existing miners to mine in a certain way, without ever needing to acquire any hardware.
This lead to a series of papers~\cite{liao2017incentivizing,velner2017smart,teutsch2016cryptocurrencies,mccorry2018smart,bonneau2018hostile} that show it is possible to introduce new incentives to an existing cryptocurrency, internally or externally, in ways that do not require trust between miners and briber (e.g., using smart contracts).
%Independent of this, Lazos et al.~\cite{lazos2019blockchain} have also proposed that miners could pay forward a reward to the first miner to extend their block, suggesting that this might incentivise honest mining behaviour.

Ethereum's uncle reward mechanism (that allows blocks that were mined but not appended to the blockchain to later be referenced in another block for a reward) can be used to subsidize the cost of bribery attacks~\cite{mccorry2018smart} and selfish mining~\cite{8406560,niu2019selfish}.
This is unfortunate, as they were originally introduced to aid decentralization~\cite{uncle} but have now been found to introduce incentives that work against this, by reducing the mining power required to perform certain attacks.

This puts into question the value of saying that a cryptocurrency is incentive compatible if new incentives can later be added.
A cryptocurrency also does not exist in a vacuum, and external incentives can always manifest in adversarial ways.
Goldfinger attacks, proposed by Kroll et al.~\cite{economicsbitcoin}, involve an adversary paying miners of a cryptocurrency to sabotage it by mining empty blocks.
In some cases, even the threat of this type of attack can be enough to kill off a cryptocurrency, as users would not want their investments to disappear if the attack happens, and thus would not invest.
As a Goldfinger attack can be implemented through a smart contract in another cryptocurrency~\cite{mccorry2018smart}, it is not inconceivable that this could be attempted in practice.
This clearly shows that incentives from outside the cryptocurrency itself must be considered.

Budish~\cite{budish2018economic} proposes an economic analysis of 51\% attack and double spending and shows that the Nakamoto consensus has inherent economic limitations.
In particular, he shows from a strictly economic point of view that the security of the blockchain relies on scarce, non-repurposable resources (i.e., ASICs) used by miners as opposed to Nakamoto's vision of ``one-CPU-one-vote'', and that the blockchain is vulnerable to sabotage at a cost linear in the amount of specialized computational equipment devoted to its maintenance.

\subsection{Other blockain consensus protocols}
As an alternative to existing systems, like Bitcoin and Ethereum, that have been shown to be vulnerable to the attacks we have just described, systems based on BlockDAGs rather than blockchains have been proposed~\cite{ghost,spectre,phantom,avalanche,fantomette,blockmania}.
In this model, the data structure is a Directed Acyclic Graph (DAG) of blocks, meaning that each block can have more than one parent block.
When creating a new block, a miner points to all the blocks that they are aware of, revealing their view of the blockchain.
This exposes more of the decision making of the players and relates to the idea of hidden actions discussed in Section~\ref{sec:dsgt}.

Due to some additional inherent flaws in Bitcoin e.g., scalability and energy consumption, new design papers are constantly proposed by both the academic community and industry, but many leave the treatment of incentives as future work.
In particular, very few papers propose an incentive scheme associated with their consensus protocols~\cite{ouroboros,fruitchains,spacemint,fantomette}.
Moreover, the solution concepts considered in these papers are often overly-simplistic; e.g., some coalition proof
NE that does not consider the impact of irrational players~\cite{ouroboros,fruitchains,snow,spacemint}.
Only Solidus~\cite{solidus} and Fant\^omette~\cite{fantomette} consider robustness (introduced in Section~\ref{sec:dsgt}).

In his draft work about incentives in Casper~\cite{caspereco}, Buterin introduces the \textit{griefing factor} which is the ratio of the penalty incurred to the victim of an attack and the penalty incurred by the attacker.
The idea of a griefing factor intuitively makes sense, as disputes in the real world can be resolved by fining a party according to the damages caused, and from a modelling point of view gives a quantifiable punishment that can be explicitly taken into account when computing equilibria.
He also proves that following the protocol in Casper is a NE as long as no player holds more than a third of the deposit at stake.

Due to the lack of formal model, it can be expected that more incentive related attacks will be proposed.
For example, attacks on cryptocurrencies using PoS are now already appearing~\cite{fanti2018compounding,stakeattack,brown2018formal}, further highlighting the need for better models.

Additionally to the consensus rules, another route to improving the
incentivization of cryptocurrencies is through their transaction fees market.
As pointed out by Lavi et al.~\cite{lavi2017redesigning}
``competition in the fee market is what keeps the rational behavior of Bitcoin's
users (partially) aligned with the goal of buying enough security for the entire system'' and is thus crucial for its security. This problem is related to that of auction theory~\cite{lavi2007computationally} and some of the literature of that field could be used here.

\subsection{Incentivizing decentralization}
Bitcoin has evolved to become different, in many ways, from the intended design and the idea of ``one-CPU-one-vote'' envisioned by Nakamoto.
Because the price of mining has increased exponentially with the popularity of Bitcoin, miners have started forming mining pools, where they join their resources
to mine, together, more blocks. %This is opposite to the decentralization envisioned in the original Bitcoin paper.
%The reason for doing so is that by pooling their resources miners will be able to mine more blocks, they will then share their gain accordingly, meaning that their average payoff is theoretically the same but their variance is reduced. 
%Due to the depreciation of the hardware this actually means that they increase their payoff by doing so.
%Mining hardware itself has changed, as ASICs have become more popular, and essential for profitable mining.

This is obviously a big threat to the security of cryptocurrencies as this could enable 51\% attacks, which have already happened to other cryptocurrencies.
As of November 2019, the most important $51\%$ attack has targeted Ethereum Classic, which is the 16th largest cryptocurrency by market cap~\cite{coinmarketcap}.

The centralization of cryptocurrencies' has been empirically analyzed by Gencer et al.~\cite{decentralizationnetwork} who measured
how decentralized Bitcoin's and Ethereum's network are.
They found that three or four mining pools control more than half of the hash power of the network.
This highlights the need for further research studying this occurrence of centralization and how decentralization can be maintained in practice.
%and try and incentivize decentralization has emerged.%~\cite{rosenfeld2011analysis}.

Several papers propose a game-theoretic analysis of the mining pools.
Arnosti et al.~\cite{arnosti2018bitcoin} model hardware investments from miners as a game,
Leonardos et al.~\cite{leonardos2019oceanic} model mining as an \emph{oceanic game}, used to analyze decision making in
settings with small numbers of big players and large numbers of individually insignificant players.
Lewenberg et al.~\cite{Lewenberg:2015:bitcoinminingpools}
% They use Cooperative Games with Coalition Structures (used by artificial intelligence researchers
% to model agent collaboration and team formation).
model the mining game as a \textit{transferable utility coalitional game}, which allows players to form coalitions and to divide their payoffs amongst themselves.
%This game is defined by a set of players and a characteristic function that specifies the monetary value that any coalition can achieve when cooperating.
%Define a miner's network.
% Define defection's function that with each coalition structure
% a set of coalitions.
As a solution concept, they use the \textit{core}, the set of feasible allocations that cannot be improved upon by a coalition, which describes stability in coalitional games.
% The core is the set of feasible allocations that cannot be improved upon by a coalition.
%same as Nash equilinirum but allows for deviation by group of agents
It captures the condition under which the agents would want to form coalitions rather than not i.e., whether there exist any sub-coalition where agents could have gained more on their own.
% A coalition is said to improve upon or block a feasible allocation if the members of that coalition are better off under another feasible allocation that is identical to the first except that every member of the coalition has a different consumption bundle that is part of an aggregate consumption bundle that can be constructed from publicly available technology and the initial endowments of each consumer in the coalition.
This concept is often opposed to the \textit{Shapley value} in Game Theory, which defines a fair way to divide the payment among the members of a coalition based on their respective contribution, but without any
consideration for stability, unlike the core.
% Transferable utility coalitional game
% cooperative games with Coalition structures
Lewenberg et al. additionally define the \textit{defection} function that captures the fact that not every agent subset can collaborate and form a new coalition.
%that associates, with each
% coalition structure (i.e., partition of the agent set
% into disjoint sets called team) a set of coalitions.
% The intuition is that the defection function captures the fact that not every agent
% subset can collaborate and form a new team.
% that, intuitively, captures the fact that not every agent
% subset can collaborate and form a new team.
%They focus on defection functions that allow for one coalition to merge
%with a subset of another coalition, or for a subset of coalition to split from its coalition.
They show that mining pools are generally unstable, no matter how the revenue is shared, some miners would be incentivized to switch to a different pool.

Eyal~\cite{minersdilemma} also studies the stability of mining pools and proposes an attack where pools infiltrate other pools to sabotage them by joining the pool and earning rewards, but without actually contributing i.e., not revealing when they find a PoW solution.
There exists configurations in which this attack constitutes a NE and an example of a tragedy of the commons.

Mining pools can also attack each other through distributed denial of service (DDoS) attacks to lower the expected success of a competing pool (large ones in particular), rather than increasing their own computational power~\cite{johnson2014game}.
Over a two year period, Vasek et al.~\cite{vasek2014empirical} found that 62.5\% of mining pools accounting for more than 5\% of the Bitcoin network power had been targeted, while only 17.1\% of the smaller pools had been targeted.
This has general implications for the mining ecosystem, as a peaceful equilibrium would require an increase to the cost of attacks and to the miner migration rate (miners switching pools), with no pool being significantly more attractive than others~\cite{laszka2015bitcoin}.

Br{\"u}njes et al.~\cite{stakepools}
introduce and study reward sharing schemes that promote the fair formation of stake pools
%in collaborative projects that involve a large number of stakeholders such as the maintenance
%of 
in a PoS blockchain. 
% Our mechanisms are parameterised by a target value for
% the desired number of pools. We show that by properly incentivising participants, the desired
% number of stake pools is a non-myopic Nash equilibrium arising from rational play.
They argue that a NE only considers \textit{myopic players}, i.e., players who ignore the responses to their own actions.
As a result, they consider the notion of \textit{non-myopic Nash equilibrium} (based on previous work by Fiat et al.~~\cite{fiat2013beyond}), which captures the effects
that a certain move will incur anticipating a strategic response from the other
players.
% Nash equilibrium the players do not
% have to take into account the impact their selection will make on the moves of the other players.
%myopic: agents act myopically and ignore responses to their own actions. 

% Other papers study the incentivization of decentralization~\cite{smartpool,scratchoff}, but without any formal model, relying instead on ad hoc arguments.
% \alex{is there a point to this sentence given that it is its own subsubsection? maybe we can just move it somewhere else since it doesn't seem very important.}
%without however formal model and while using "ad hoc" arguments.

Luu et al.~\cite{smartpool} use smart contracts to decentralize mining by
incuring mining fees lower than centralized mining pools.
Miller et al.~\cite{scratchoff} present several definitions and constructions
for ``non-outsourceable'' puzzles. Both  papers use informal arguments to justify their
construction as opposed to a formal model.

% This once more highlights the trend of attack-then-patch and is another
% argument for having a strong incentive model that deal with (de)centralization.
% \alex{Just reads weird after the last subsubsection that seems inconsequential.}

In a recent paper, Kwon et al.~\cite{kwon2019impossibility} propose a formal model
for the decentralization of blockchains, and show that full
decentralization is impossible unless there exists a Sybil cost.

%\subsubsection{Other open problems}
\subsection{Payment Channels}
In order to overcome the scalability issues of Bitcoin,
a new concept, referred to as \textit{layer 2} or \textit{payment channels} has been
proposed~\cite{towards}.
The idea is that since the network cannot handle enough transactions, participants
can take some transactions off-chain i.e., outside the main network, by opening a channel between themselves. % and exchanging money however they please.
This is done by locking a deposit on the blockchain, opening the channel and transacting on the channel, then settling the overall balance of all transactions on-chain so that the blockchain will see only two transactions (locking the funds and settling the balance).
%and then only settle online (on-chain) when they decide to stop transacting.
% The blockchain will see only two transactions, the one that locks funds and the one
% that settles the final balances, greatly reducing the amount of on-chain transactions.

Several designs have been proposed to achieve this~\cite{towards,sprites,perun}.
The high level idea is that participants will create evidence of each of their transactions (e.g., using signatures) so that whenever someone tries to cheat the other party can prove it and receive the cheating party's deposit as compensation.
% As as a compensation, they will get all the deposited funds of the cheating party.
%For example, if Alice pays 1 bitcoin to Bob who then pays 2 bitcoin to Alice, Bob
%could try to cheat by broadcasting the first transactions to the blockchain, with the obsolete
%balances, but Alice could broadcast to the blockchain the transaction signed by Bob to prove
%cheating has occurred.

In this setting, the security relies on the fact that cheating is easily
detectable due to cryptographic evidence and on the financial
punishment associated with it. So again, incentives are tightly
linked to security.
A few papers~\cite{sprites,statechannels} present formal models to analyze the security of these payment channels.
They are based on the UC-model mentioned in Section~\ref{sec:prelim} but do not consider utilities although it is an important part of the security of the system.

In order to facilitate payment channels, a routing solution has been proposed~\cite{lightning-network}.
%The idea is that if Alice wants to open a channel with Bob and both Bob and Alice already have an existing channel open with Charlie, then Charlie can act as a router between Alice and Bob, without them needing to open a new channel.
There are usually difficulties in this case, due to the need for collaterals to be locked by everyone on the routing path.
This work is related to the one on selfish routing discussed in
Section~\ref{sec:dsgt}.

A problem with payment channels is the requirement for participants to be online to detect cheating i.e., the cheater broadcasting an old balance to the blockchain.
McCorry et al.~\cite{pisa} propose delegating this task to a third party, a \emph{watch tower}, but it is unclear how incentives should be designed in this context.

\subsection{Fairness}
% \saraha{find that paper by Joe about random beacon on bitcoin}
% \url{https://eprint.iacr.org/2015/1015.pdf}
% \url{https://github.com/Mechanism-Labs/MetaAnalysis-of-Alternative-Consensus-Protocols/blob/master/MetaAnalysis.pdf} A metaAnalysis of proposed alternative consensus protocols for blockchains
Fairness in cryptocurrencies is implicitely captured by the notion of chain quality introduced by Garay et al.~\cite{bitcoin-backbone}, which states that an adversary should not contribute
more blocks to the blockchain than what they are supposed to i.e., proportionally
to their computational power in the PoW setting.

Chen et al.~\cite{chen2019axiomatic} recently showed that a proportional reward system is the unique allocation rule that satisfies properties of symmetry, budget balance (weak or strong), sybil-proofness, and collusion-proofness, which are desirable.

In the PoS setting, Fanti et al.~\cite{fanti2018compounding} define the notion of \textit{equitability} that corresponds to how much a node's initial investment (i.e., stake) can grow or shrink, and address the problem of the ``rich get richer'' in PoS cryptocurrencies (which arguably also exists in PoW cryptocurrencies).
They propose a geometric reward function that they prove is more equitable i.e., the distribution of stake stays more stable.
In general, the problem of compounding of wealth is reinforced by the fact that early adopters of a cryptocurrency have a significant advantage, benefiting from the ease of mining (or staking) and greatly cheaper coin prices in the early days.
Dealing with this is more of a macroeconomic problem that to the best of our knowledge has not yet received any attention.

\section{Discussion}\label{sec:discussion}
Within each section of this paper, we have reviewed models that draw on game theoretic tools and could thus be used to address problems in blockchain based cryptocurrencies.
Table~\ref{table:discussion} gives a summary of some these models, and we will now discuss general points and lessons that can be drawn from the work so far.

\begin{table*}[t!]
\large
\begin{center}
\caption{Summary of the models surveyed in this work, along with the problem they seek to address and (if applicable) shortcomings. }
\label{table:discussion}
\resizebox{\linewidth}{!} {%
\begin{tabular} {p{3.5cm}  p{3.5cm}  p{6.5cm} p{3.5cm}  p{6.5cm}}
\toprule 
\textbf{Concept} & \textbf{Player types} & \textbf{Description} & \textbf{Shortcomings} & \textbf{Example, suggestions, and discussion}\\
\midrule 
RPD & Rational, honest & Meta-game between the game designer and the adversary, which the adversary wins if it can exploit a vulnerabilities without decreasing its utility. & Does not consider irrational adversaries. & Could be used in Layer 2 since in that case a misbehaving participants will be penalized and the other participant will earn all of its money so Byzantine adversaries cannot harm honest players more than rational adversaries.\\
\midrule 
BMG & Rational & Considers games and the complexity of actions by modelling players as Turing machines & Results are primarily equivalence theorems with dense proofs rather than new results or simpler proof methods. No consideration of Byzantine adversaries. & This could be usefully applied to cryptocurrencies given that many situations are modeled as games involving computation e.g., mining.\\
\midrule
Price of Unaccountability & Rational &  Worst ratio between utilities in the observable-actions case and the hidden actions case. & Defined in the principal-agent model. & As differentiating between malicious behavior and genuine latency is hard, especially in PoS systems, the Price of Unaccountability may be a useful to evaluating this. \\
\midrule 
BAR & Byzantine, rational, honest & Introduces the idea of BART and IC-BFT protocols, which make it possible to bypass impossibility results in consensus protocols.  &  & \\
\midrule
$(k,t)$-robustness & Byzantine, rational, honest & Extension of NE that consider coalitions of both rational and Byzantine players. & The two concepts of immunity and robustness are treated separetely. & This model is better suited than NE to study the game theoretic aspects of consensus protocols that involve many players and coalitions.\\
\midrule 
Price of Malice and Byzantine Price of Anarchy & Byzantine, rational & Quantifies how much a system degrades with the presence of irrational players and relates to the concept of immunity introduced in $(k,t)$-robutsness. &  Does not capture how different information sets impact the system. & Could inspire new measures that quantify the trade-off between blockchain and traditional consensus protocols. Blockchain-based systems are intended to be more scalable as they are meant to handle open participation, compared to classical consensus that requires the many messages to be exchanged, but in the case of Bitcoin this comes at the price of PoW so there is an incurred economic cost. \\
\midrule 
% Byzantine Price of Anarchy & Rational and Byzantine & This quantifies how much an optimal system degrades due to selfish behavior, when malicious players are introduced. & & \\

Fear factor & Byzantine, rational & Rational players are incentivized to follow the protocol by fear of Byzantine players. Similar to Price of Malice (if the system improves instead of deteriorates). & Same as above. & This could be used to argue against the verifier dilemma. Some users may be motivated to verify the content of blocks by fear that others will not.\\

% revelation principle & Rational &  & & It has been argued to have been used in system like bitcoin-NG, fruitchain, blockdag etc as they ``reveal'' the rational strategy \saraha{add more}\\
\bottomrule
\end{tabular}
}
\end{center}
\end{table*}

The most immediate problem that most models try to address is the consideration of incentives at the level of security properties by encoding utility functions and some notion of equilibrium into a modified definition of the traditional security property.
For example, RPD does this for UC security and the BAR model does this for distributed systems. 
Already, however, some differences emerge.

First, there are differences in the types of players. 
Different models assume players can be some subset of honest, rational, or Byzantine.
How well the assumed types of players reflect the reality of the system will have a great impact on the usefulness of the model, regardless of its technical merits.

In particular, while honest and Byzantine players can be defined with respect to a protocol and whether or not they follow it, the meaning of a rational player is harder to pin down. 
This is because a player is in practice not rational only with respect to a protocol and the incentives in the system that the protocol is part of, but also with respect to a variety of possible external incentives.
We have referred to this with respect to rational cryptography and bribery attacks, but it is a more general point that Ford and B{\"o}hme have explicitely highlighted~\cite{ford2019rationality}.

Security models are built such that within the model, properties can remain true up to some amount of adversarial capabilities e.g., a third of Byzantine nodes and computational capabilities, but it is not clear how valid the models are when the assumed player types differ from reality.
Moreover, comparing the models presented in this paper would require a concrete idea of what the correct assumptions that can be made about participants in cryptocurrencies are.
Each model is constructed to work better than others within the constraints of their assumptions.
As remarked by Box, ``all models are wrong, but some are useful''~\cite{box1976science}.
It seems that the current limitation of the exisiting litterature is in understanding how useful these models are in practice rather than in introducing new models.

In Economics, Becker pointed out long ago that it was not always clear what rationality implied, because some observed behaviour was compatible with both rational and irrational behaviour~\cite{10.2307/1827018}, and concluded that perhaps irrationality deserved to be studied with more attention.
Behavioural Economics and, more broadly, experimental data driven work has taken an important place in Economics to understand why certain models did not work in practice.
Perhaps the same should be done here.

For example, mining rewards can be designed as part of a protocol to ensure some level of decentralization but the utility function of miners will also depend on their individual economic environment that the protocol cannot fix. 
In this case, relating the economics of miners to their relationship with the system could do more to ensure some level of decentralization than tweaking the rewards given out by the protocol.

% This is also true regarding the choice of equilibrium
% Fundamentally, comparing the models presented in this paper would require a concrete idea of what the correct assumptions that can be made about participants in cryptocurrencies are.
% Each model is constructed to work better than others within the constraints of their assumptions.

Second, there are the way coalitions of players are treated.
Pools are being observed and studied carefully in the community, but some important problems remain under-studied.
It could be argued that in some cases pools somehow self-regulate as in, for example, the case of the Ghash.io Bitcoin pool that once got more than 51\% of the hashing power
but then decided to withdraw part of it~\cite{ghash}.
This could be because of the fear of loss of confidence users of the system, which if it is justified, would mean that the incentives were to some extent aligned to prevent a 51\% attack.

Coalitions are also easily abstracted as an entity controlling a fixed share of power in the system but this ignores costs inherent to colluding such as, for example, communications costs as a coalition much reach some form of consensus on what actions it takes.
In the same way that results are obtained for a whole system based on assumed proportions of player types in the system, perhaps useful results can be obtained based on the proportions of player types in a coalition with respect to that coallition.
% Same goes for attacks such as selfish mining that are not being widely observed.

% Finally, many of the models here opt to be as formal as possible, sometimes to the point of relying on results that may not be applicable in practice. 
% Computing equilibriais an important part of algorithmic Game Theory due to the inherent difficulty in computing them.
% In particular, Daskalkis et al.~\cite{daskalakis2009complexity} pointed out

%\saraha{argue NE and utilities function are hard to calculate}

%\input{table}
%!TEX root = main.tex
\section{Conclusion}
%People keep reinventing the same concepts. Hopefully this paper will help.
Security researchers and cryptographers have been interested in incorporating
game theoretic notions to their models for many years.
In this work, we have highlighted existing concepts and explained how and where they could be used for
specific applications.

The approach taken in most of the papers that we described here is to extend
a field by for example incorporating utility functions (Rational Cryptography)
or computation (Bayesian Machine Games).
No completely new theory has appeared and it
would be interesting to see a new theory built from the ground up to address considerations of incentives at all stages of the design process, rather than adapting existing models.
We hope that this paper will give some inspiration towards new formal models.
% create new formal concepts that derive from
% those presented here.

% \section{acks}
% Alexander Hicks is supported by OneSpan\footnote{https://www.onespan.com} and UCL through an EPSRC Research Studentship.

% {\footnotesize
% \def\shortbib{1}
% \bibliographystyle{abbrv}
% \bibliography{../abbrev2,../crypto,../misc,alex}
% \bibliography{../crypto,../misc,alex}
\bibliographystyle{plain}
\bibliography{../crypto,../misc,alex}

\appendix
%\input{app-papers}
%\input{prelim}
%!TEX root = main.tex
\section{Glossary}
\label{glossary}
In this appendix, we provide formal definitions for some of the concepts presented in the main body of the paper that are not formally defined.

\subsection{Game Theory}
To start off, we introduce the standard definitions for Bayesian games and mechanisms.
\begin{description}
\item[Bayesian game setting]
A Bayesian game setting is a tuple $(N,O,\Theta,\Pr,u)$, where:
\begin{itemize}
\item N is a finite set of $n$ players;
\item O is a set of outcomes;
\item $\Theta = \Theta_1,\cdots,\Theta_n$ is a set of possible joint type vectors
\item $\Pr$ is a (common prior) probability distribution on $\Theta$; and
\item $u=(u_1,\cdots,u_n)$, where $u_i: O\times \mathbb{R} \rightarrow \mathbb{R}$ is the utility
function for each player $i$
\end{itemize}
%Given a Bayesian game, we define as Mechanism as follows:

\item[Mechanism for a Bayesian game setting]
A mechanism for a Bayesian game setting $(N,O,\Theta,p,u)$ is a pair (A,M), where
\begin{itemize}
\item $A=A_1\times\cdots\times A_n$, where $A_i$ is the set of actions available to agent $i\in N$
\item $M:A\rightarrow \mathcal{D}(O)$ maps each action profile to a distribution over outcomes
\end{itemize}

% \item[Nash Equilibrium]
% \item[ex-post NE]
\end{description}

\subsection{Game Theory and Cryptography}
We now move on to concepts presented in Section~\ref{sec:cryptoandgt}.
\begin{description}
%\item[Covert Adversary]~\cite{aumann2007security}
\item[$\epsilon-$subgame perfect equilibrium~\cite{rationalcrypto}]
Let $\mathcal{G}_{\mathcal{M}}$ be an attack game. A staretgy profile $(A,\Pi)$
is an $\epsilon-$subgame perfect equilibrium in $\mathcal{G}_{\mathcal{M}}$ if:
(1) for any $\Pi'\in \trm{ITM}^n$, $u_D(\Pi',A(\Pi'))\le u_D(\Pi,A(\Pi))+\epsilon$,
and (2) for any $A'\in \trm{ITM}$, $u_A(\Pi,A'(\Pi))\le u_A(\Pi,A(\Pi)) +\epsilon$.

\item[Attack-payoff security~\cite{rationalcrypto}]
Let $\mathcal{M}=(\mathcal{F},\langle\mathcal{F}\rangle,v)$ be an attack model and let
$\Pi$ be a protocol that realizes functionality $\langle\mathcal{F}\rangle$.
$\Pi$ is attack-payoff secure in $\mathcal{M}$ if $\vec{U}^{\Pi,\langle\mathcal{F}\rangle}
\overset{\trm{negl}}{\le}\vec{U}^{\Phi^\mathcal{F},\langle\mathcal{F}\rangle}$
where $\Phi^\mathcal{F}$ is the ``dummy'' $\mathcal{F}$ hybrid protocol (i.e.,the protocol
that forwards all inputs and outputs from the functionality $\mathcal{F}$,
see Section~\ref{sec:prelim}) and $\vec{U}^{\Pi,\langle\mathcal{F}\rangle}$ is the maximized ideal expected payoff of an adversary.

\item[Incentive compatibility~\cite{butwhy}]
Let $\Pi$ be a protocol and $\mathds{P}$ be a set of PT protocols that have access to
the same hybrids as $\Pi$. We say that $\Pi$ is $\mathds{P}-$ incentive compatible in
the attack model $\mathcal{M}$ if and only if for some $\adv$ $(\Pi,\adv)$ is a $(\mathds{P},\trm{ITM})-$
subgame perfect equilibrium in the attack game defined by $\mathcal{M}$.

\item[Bayesian Machine Game~\cite{2008arXiv0809.0024H}]
A Bayesian machine game G is described by a tuple $(N, \mathcal{M},\Theta,\Pr,\mathcal{C}_1,\dots,\mathcal{C}_m,u_1,\dots,u_2)$ where:
\begin{itemize}
	\item $N$ is the set of players, $\mathcal{M}$ is the set of possible machines
	\item $\Theta\subseteq(\{0,1\}^*)^{m+1}$ is the set of type profiles where the $(m+1)$st element in the profile corresponds to nature's type
	\item $\Pr$ is a distribution on $\Theta$
	\item $\mathcal{C}_i$ is a complexity function
	\item $u_i:T\times(\{0,1\}^*)^m\times\mathbb{N}\rightarrow\mathbb{R}$ is player i's utility function.
\end{itemize}
Given a Bayesian machine game G, a machine profile $\vec{M}$, and $\epsilon\geq0$, $M_i$ is an $\epsilon$-best response to $\vec{M}_{-i}$ (the tuple consisting of all machines in $\vec{M}$ other than $M_i$) if, for every $M^{'}_{i}\in\mathcal{M}$,
\begin{equation}
U^G_i[(M_{i},\vec{M}_{-i})]\geq U^G_i[(M^{'}_{i},\vec{M}_{-i})]-\epsilon.
\end{equation}
$\vec{M}$ is an $\epsilon$-Nash equilibrium of G if, for all players i, $M_i$ is an $\epsilon$-best response to $\vec{M}_{-i}$. A Nash equilibrium is a $0$-Nash equilibrium.

\item[Universal implementation~\cite{2008arXiv0809.0024H}]
Suppose that $\mathcal{G}$ is a set of n-player canonical games, $\mathcal{Z}$
is a subsets of $N$, $\mathcal{F}$ and $\mathcal{F}'$ are mediators,
$M_1,\cdots,M_n$ are interactive machines, $p:\mathbb{N}\times\mathbb{N}\rightarrow\mathbb{N}$
and $\epsilon:\mathbb{N}\rightarrow\mathbb{R}$.
$(M,\mathcal{F}'$ is a $(\mathcal{G},\mathcal{Z},p)$-universal implementation
of $\mathcal{F}$ with error $\epsilon$ if, for all $n$, all games $G\in\mathcal{G}$
with input length n and all $\mathcal{Z}'\subseteq\mathcal{Z}$
if $\vec{\Lambda}^\mathcal{F}$ is a $p(n,\cdot)$-robust $\mathcal{Z}'$-safe $\epsilon$-NE
in the mediated machine game $(G,\mathcal{F})$ then
\begin{enumerate}
\item (Preserving equilibrium) $\vec{M}$ is a $\mathcal{Z}'$-safe $\epsilon$-NE
in the mediated machine game $(G,\mathcal{F}')$
\item (Preserving Action Distributions) For each type profile $\vec{t}$,
the action profile induced by $\vec{\Lambda}^\mathcal{F}$ in $(G,\mathcal{F})$
is identically distributed to then action profile induced by M in $(G,\mathcal{F}')$.
\end{enumerate}

\item[Sequential equilibrium in computational games~\cite{halpern2013sequential}]
A pair $(\vec{M},\mu)$ consisting of a machine profile $\vec{M}$ and a belief system $\mu$ is called a belief assessment.
A belief assessment $(\vec{M},\mu)$ is an interim (resp. ex ante) sequential equilibrium in a machine game $G$ if $\mu$ is compatible with $\vec{M}$ and for all players $i$, states $q$ of $M_i$, and machines $M_i^{'}$ compatible with $M_i$ and $q$ such that $(M_i,q,M_i^{'})\in\mathcal{M}$ (the set of possible machines) (resp. $(M_i,q,M_i^{'})$ is a local variant of $M_i$), we have
\begin{equation}
U_i(\vec{M}\vert q,\mu) \geq U_i(((M_i,q,M_i^{'}),\vec{M}_{-i})\vert q, \mu)
\end{equation}
\end{description}

\subsection{Game Theory and Distributed Design}
Finally, we give definitions for concepts presented in Section~\ref{sec:dsgt}.
\begin{description}
\item[Incentive-Compatible Byzantine Fault Tolerant (IC-BFT) protocols~\cite{barmodel}]
A protocol is IC-BFT if it guarantees the specified set
of safety and liveness properties and if it is in the best interest
of all rational nodes to follow the protocol exactly.
\item[Byzantine Altruistic Rational Tolerant (BART) protocols~\cite{barmodel}]
 A
protocol is BART if it guarantees the specified set of safety
and liveness properties in the presence of all rational deviations
from the protocol.

\item[Perfect security~\cite{barational} ]
A protocol for broadcast or consensus is perfectly
secure against rational adversaries controlling t players with utility U if
for every t-adversary there is a strategy S such that for any choice of input for
honest players
1. (S is tolerable): S induces a distribution of final outputs D in which no
security condition is violated with nonzero probability, and
2. (S is Nash): For any strategy $S'\ne S$ with induced output distribution D'
:
$U(D)\ge U(D')$.

\item[Statistical Security~\cite{barational}]
A protocol for broadcast or consensus is
statistically secure against rational adversaries controlling t players with utility
U if for every t-adversary there is a strategy S such that for any choice of input
for honest players S induces a distribution of final outputs $D_k$ when the security
parameter is k and the following properties hold:
1. (S is tolerable): no security condition is violated with nonzero probability in
Dk for any k, and
2. (S is statistical Nash): for any strategy $S'
\ne S$ with induced output distributions
$D'_k$
there is a negligible function $negl(.)$ such that $U(D_k) + negl(k) >
U(D'_k)$.

\item[(k,t)-robustness~\cite{multiparty-computation-game-theory}]
A strategy profile $\sigma$ is a (k,t)-robust equilibrium if
for all $C,T\subseteq N,\ C\cap T =\emptyset,\ |C|\le k,\ |T|\le t$
$\forall \tau_t\in\strategyset_T\ \forall \phi_C \in C$ we have:
$u_i(\strategy_{-T},\tau_T)\ge u_i(\strategy_{-C\cap T},\phi_C,\tau_T)$

\item[(k,t)-punishment~\cite{multiparty-computation-game-theory}]A joint strategy $\rho$ is a (k, t)-punishment
strategy with respect to $\sigma$ if for all $C, T, P \subseteq N $ such that
C, T, P are disjoint, $|C|\le k,\ |T| \le t,\ \textrm{and}\ |P| > t$, for all
$\tau_T \in S_T$ , for all $ \phi_C \in S_C$ , for all $i \in C$ we have
$u_i(\sigma_T , \tau_T ) > u_i(\sigma_{N-(C\cup T \cup P )}, \phi_C , \tau_T , \rho_P )$.

%\item[BAR Model]
% \item[Hidden actions]
% \item[Direct versus Indirect Reciprocity]
% \item[Price of Malice]
% \item[Price of Unaccountability]
% %%% Information related concepts
% \item[Oblivious model]~\cite{selfishmeetsevil} selfish players are aware of the existence
% of Byzantine players

%%solution concepts (how players behave)

% \item[Cooperative games with  coalition structures]~\cite{Lewenberg:2015:bitcoinminingpools}
% \item[transferable utility coalitional game]
% \item[Core]~\cite{Lewenberg:2015:bitcoinminingpools}
% \item[Shapley Value]
% \item[Defection function]~\cite{Lewenberg:2015:bitcoinminingpools}
%%security properties

% \item[Non-myopic Nash Equilibrium]

%%Measures

% \item[Defection function]~\cite{Lewenberg:2015:bitcoinminingpools}
% \item[Characteristic function]~\cite{Lewenberg:2015:bitcoinminingpools}specifies the monetary value that any coalition can achieve when cooperating.

\end{description}

\end{document}